\documentclass[conference]{IEEEtran}
\IEEEoverridecommandlockouts
\usepackage{cite}
\usepackage{amsmath,amssymb,amsfonts}
\usepackage{algorithmic}
\usepackage{graphicx}
\usepackage{textcomp}
\usepackage{xcolor}
\usepackage{algorithm}
\usepackage{multirow}
\usepackage{makecell}
\usepackage{multirow}
\usepackage{bbding}
\usepackage{array}
\usepackage{bm}  
\usepackage{float}
\usepackage{stfloats}
\usepackage[a4paper, total={184mm,239mm}]{geometry}
\def\BibTeX{{\rm B\kern-.05em{\sc i\kern-.025em b}\kern-.08em
    T\kern-.1667em\lower.7ex\hbox{E}\kern-.125emX}}
    \DeclareRobustCommand*{\IEEEauthorrefmark}[1]{%
    \raisebox{0pt}[0pt][0pt]{\textsuperscript{\footnotesize\ensuremath{#1}}}}

\begin{document}

\title{AmpAgent: An LLM-based Multi-Agent System for Multi-stage Amplifier Schematic Design from Literature for Process and Performance Porting
}

\author{
\IEEEauthorblockN{
Chengjie Liu\IEEEauthorrefmark{1,3},
Weiyu Chen\IEEEauthorrefmark{1,3},
Anlan Peng\IEEEauthorrefmark{1,3},
Yuan Du\IEEEauthorrefmark{1}
Li Du\IEEEauthorrefmark{1},and
Jun Yang\IEEEauthorrefmark{2,3}}
\IEEEauthorblockA{\IEEEauthorrefmark{1}School of Electronic Science and Engineering, Nanjing University, Nanjing, China}
\IEEEauthorblockA{\IEEEauthorrefmark{2}School of Integrated Circuit, South East University, Nanjing, China}
\IEEEauthorblockA{\IEEEauthorrefmark{3}National Center of Technology
Innovation for EDA, Nanjing, China}
\IEEEauthorblockA{cjliu\_phd@smail.nju.edu.cn, wychen@smail.nju.edu.cn, alpeng@smail.nju.edu.cn}
\IEEEauthorblockA{yuandu@nju.edu.cn, ldu@nju.edu.cn, dragon@seu.edu.cn}
\IEEEauthorblockA{Corresponding Author: Li Du \quad Email: ldu@nju.edu.cn}
}

\maketitle

\begin{abstract}
Multi-stage amplifiers are widely applied in analog circuits. However, their large number of components, complex transfer functions, and intricate pole-zero distributions necessitate extensive manpower for derivation and device parameters sizing to ensure their stability.
In order to achieve efficient transfer function derivation and device parameters sizing, thereby simplifying the difficulty of amplifier design, we propose AmpAgent: a multi-agent system based on large language models (LLMs) for efficiently designing such complex amplifiers from literature with process and performance porting. AmpAgent is composed of three agents: Literature Analysis Agent, Mathematics Reasoning Agent and Device Sizing Agent. They are separately responsible for retrieving key information (e.g. formulas and transfer functions) from the literature, decompose the whole circuit's design problem by deriving the key formulas, and address the decomposed problem iteratively.

AmpAgent was employed in the schematic design of seven types of multi-stage amplifiers with different compensation techniques. 
In terms of design efficiency, AmpAgent has reduced the number of iterations by 1.32$ \sim $4${\times}$ and execution time by 1.19$ \sim $2.99${\times}$ compared to conventional optimization algorithms, with a success rate increased by 1.03$ \sim $6.79${\times}$. In terms of circuit performance, it has improved by 1.63$ \sim $27.25${\times}$ compared to the original literature.
The findings suggest that LLMs could play a crucial role in the field of complex analog circuit schematic design, as well as process and performance porting.

\end{abstract}


\section{Introduction}

Multi-stage amplifier, compared with single-stage amplifier, can provide high open-loop gain and strong driving capability but with relative low supply voltage requirement, making it useful for amplifying small signals, driving LEDs, and processing biological signals \cite{you1997multistage,948432,allen2011cmos,leung1999damping,leung1999right,peng2005transconductance,peng2010impedance,yan20130}. Designing multi-stage amplifiers, despite following a well-defined flow \cite{cannizzaro2007design}, remains challenging due to the large number of device components, complex transfer functions, and intricate pole-zero distributions compared to single-stage amplifiers. These factors not only make it challenging to manually design multi-stage amplifiers across various processes and performance metrics, but the vast search space also poses a significant challenge for optimization algorithms to design effectively.

The inefficiency in the automated design of multi-stage amplifiers lies in the inability to decompose their design problems into optimization problems at each stage. Recently, the emerging large language models (LLMs) have been proven capable of solving similar reasoning problems encountered in other fields\cite{wen2023dilu,wang2023voyager}, even conducting a set of Nobel-Prize-winning chemistry experiments\cite{boiko2023autonomous}. These demonstrate that LLMs also possess the potential to implement solutions for the automatic design of not only digital circuits \cite{blocklove2023chip,thakur2023benchmarking,liu2023chipnemo} but also analog circuits like multi-stage amplifiers.

There has been some works \cite{artisan,liu2024ladac,yin2024ado,lai2024analogcoder} that utilize LLMs to enable automated schematic design of analog circuits.
The authors of \cite{artisan} have developed an operational-amplifier-domain LLM Artisan to autonomously generate custom amplifier schematics.
The researchers presented in \cite{liu2024ladac} have developed a LLM-based agent, LADAC, and successfully automates the design process for three types of analog circuits.
Another work \cite{yin2024ado} merged the LLM agent with Bayesian Optimization (BO) algorithm \cite{NEURIPS2019_6c990b7a}, to design a 2-stage amplifier and a hysteresis comparator with greater efficiency than that of BO. AnalogCoder \cite{lai2024analogcoder} achieved designing analog circuits through Python code generation with greater capability than the state-of-the-art (SOTA) LLMs.

Despite the above work demonstrating the capability of LLMs in designing analog circuits effectively, there are still unresolved challenges in designing complex multi-stage amplifiers. For example, a Nested Gm-C Compensation amplifier\cite{you1997multistage}, which is composed of 18 components, has 2 compensation caps and 2 feedforward stage to stabilize the amplifier which is hard for current LLMs to design due to the following reasons:
\begin{itemize}
    \item The lack of analog circuits data results in even the SOTA LLMs lacking the relevant knowledge to provide valuable design recommendations for multi-stage amplifiers;
    \item Despite being provided with detailed amplifier  information, LLMs' ability to utilize this information for amplifier design is hindered by a lack of understanding of their design process.
\end{itemize}

To address the aforementioned issues, we developed AmpAgent, an LLM-based multi-agent system, for multi-stage amplifier schematic design from literature with process and performance porting instead of just replication\cite{xiong2024ai}.
AmpAgent incorporates the following three pipeline agents to address the two issues mentioned above:
\begin{itemize}
    \item Literature Analysis Agent: Used for extracting essential design details on multi-stage amplifiers from literature, supplementing the current LLMs' knowledge gaps and ensuring precise data for downstream applications.
    \item Mathematics Reasoning Agent: Responsible for deducing important formulas in conjunction with information such as the stability conditions of amplifiers. This decomposes the overall optimization problem of the amplifier into several sub-problems, thereby leveraging the results from upstream;
    \item Device Sizing Agent: Integrated with simulation interface and conventional optimization algorithms to size the multi-stage amplifier according to the various sub-problems divided by the previous agent efficiently.
\end{itemize}

Through the collaboration of the three pipeline agents, AmpAgent successfully achieved rapid schematic design of multi-stage amplifiers from literature.
The primary contributions of this paper are:
\begin{itemize}
    \item A novel domain-specified LLM multi-agent system is developed to realize the schematic design of multi-stage amplifier according to literature, incorporating process and performance porting.
    \item The performance of the AmpAgent-designed multi-stage amplifier has an enhancement of 27.25$\times$ compared to the literature's;
    \item Compared to the conventional optimization algorithms, achieving a reduction of up to 4${\times}$ in the number of algorithm iterations, an execution time reduction of up to 2.99${\times}$, and an increase in success rate of up to 6.79${\times}$.
    
\end{itemize}

In summary, our work demonstrates the potential of LLMs in the analog circuits' schematic design, as well as process and performance porting. This paper is organized as follows: 
Section II provides a detailed discussion on the current limitations of LLMs in analog circuits.
Section III outlines the overall workflow of AmpAgent.
Section IV exhibits the experiment of applying AmpAgent to the schematic design of seven different kinds of multi-stage amplifiers and AmpAgent's design efficiency compared to conventional optimization algorithm.
Finally, Section V concludes by summarizing our work.

\section{LLM for analog circuits}
In this section, we will demonstrate the necessity of constructing the multi-agent system within AmpAgent by showcasing the inadequacies of SOTA large models, including GPT-3.5, GPT-4\cite{OpenAI2023GPT4TR}, and GPT-4o\cite{openaigpt-4o}, in dealing with analog circuits.

The study presented in \cite{liu2024ladac} explored the application of GPT-4 in generating SPICE netlists for a $3^{rd}$-order low-pass RC filter and a five-transistor amplifier. The results indicated that GPT-4 was unsuccessful in this task. The research in \cite{yin2024ado} integrated GPT-3.5 with BO to enhance the efficiency of analog circuit sizing. Despite this, reliance on GPT-3.5 alone led to a substantially higher number of iterations before finalizing a circuit design.
\begin{figure}[!t]
  \centering
  \includegraphics[width=3.3in]{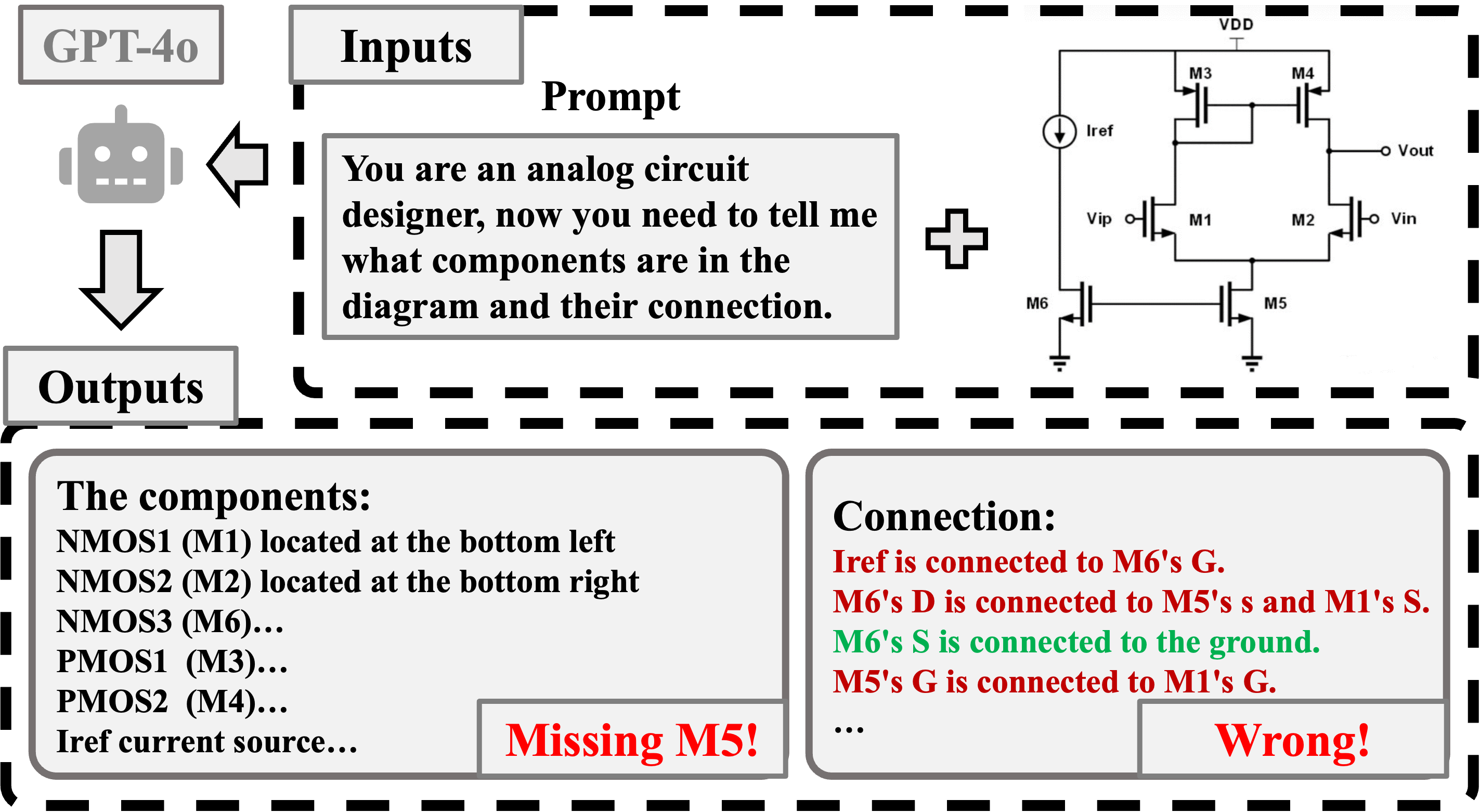}
  \caption{Experiment on GPT-4o extracting netlist from circuit diagram}
  \label{gpt_4o_test}
\end{figure}

We further evaluated GPT-4o's\cite{openaigpt-4o} ability to extract the netlist from the circuit diagram, as shown in Fig.\ref{gpt_4o_test}. Unfortunately, GPT-4o fails to accurately identify the components and their corresponding connections.

The experiment results of previous works \cite{liu2024ladac,yin2024ado} and the ability of GPT-4o\cite{openaigpt-4o} to recognize analog circuit diagrams will be presented. These experiments, which is illustrated in Table \ref{tab_gpt_capability}, have shown that even the current SOTA LLMs lack sufficient knowledge about analog circuits, thereby also lacking the capability to accurately recognize circuits.
Therefore, it is necessary to utilize a multi-agent system to establish division and collaboration in the design of analog circuits, thereby reducing the capability requirements for each agent and enhancing the performance of LLMs in designing circuits.

\renewcommand{\arraystretch}{1}
\begin{table}
    \begin{center}
    \caption{GPT-4's and GPT-4o's capacity for analog circuit design}
    \label{tab_gpt_capability}
    \begin{tabular}{|c|c|c|}
    \hline
    Model Name & Experiment & Result \\
    \hline
    GPT-4\cite{OpenAI2023GPT4TR} & Spice generation\cite{liu2024ladac} & \XSolidBrush \\
    \hline
    \multirow{2}{*}{GPT-3.5}& Sizing Suggestion (Rough)\cite{yin2024ado}  & \CheckmarkBold \\
    \cline{2-3}
    & Sizing Suggestion (Precise)\cite{yin2024ado}  & \XSolidBrush \\ 
    \hline
    GPT-4o\cite{openaigpt-4o} & Circuit Diagram Recognition & \XSolidBrush \\
    \hline
    \end{tabular}
    \end{center}
\end{table}

\begin{figure*}[!t]
\centering 
\includegraphics[width=6.8in]{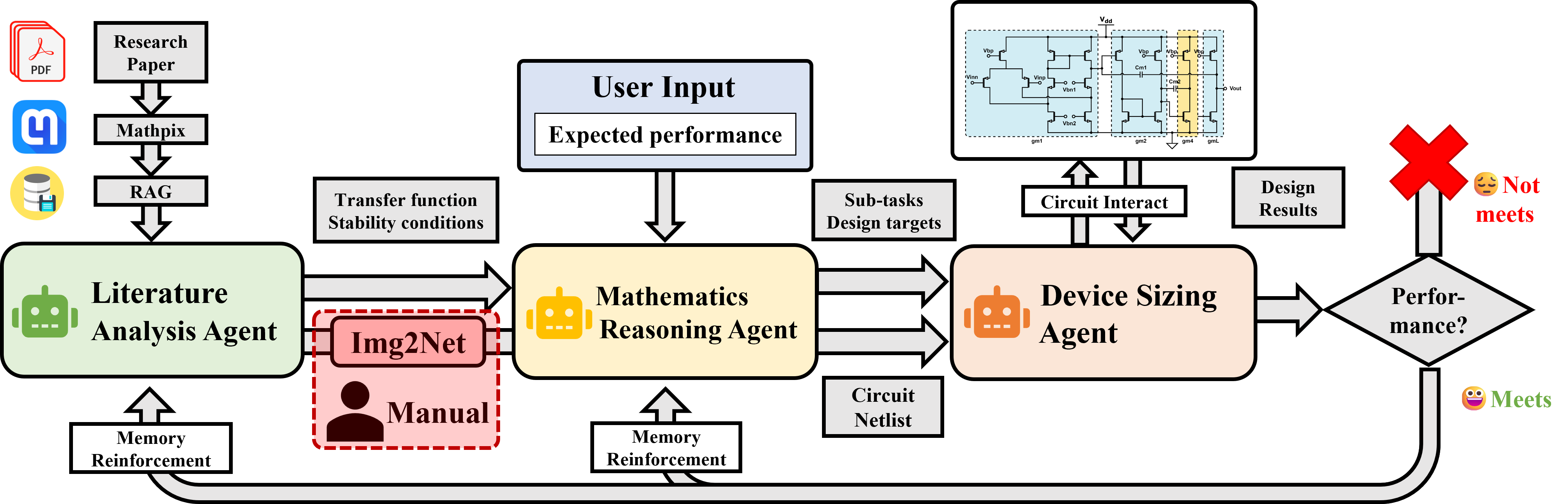} 
\caption{The overview of AmpAgent} 
\label{Fig3} 
\end{figure*}
\section{Overview of AmpAgent}
AmpAgent is an LLM-based multi-agent system designed for designing multi-stage amplifier schematics from literature. The workflow of AmpAgent is shown in Fig. \ref{Fig3}. AmpAgent comprises three agents: (1) Literature Analysis Agent, (2) Mathematics Reasoning Agent, and (3) Device Sizing Agent. All agents are developed based on ReAct \cite{yao2022react} technique.

The Literature Analysis Agent is capable of retrieving and summarizing key information such as critical computational formulas and stability conditions. This is realized by embedding the under-designed literature as the vector database of the Literature Analysis Agent's RAG.
 
The Mathematics Reasoning Agent, in conjunction with the user's input of expected performance and the key formulas retrieved and summarized by the Literature Analysis Agent, carries out predictive calculations for the critical parameters of multi-stage amplifiers, such as the transconductance ($g_m$) of each stage. Thereby, it decomposes the overall design problem into multiple sub-problems.

The Device Sizing Agent interacts with the amplifier (such as sizing the devices and reading simulation results) and, in conjunction with conventional optimization algorithms, sequentially resolves the sub-problems.

In addition to the three agents mentioned above, the intermediate reasoning steps will be saved after confirming the correct functionality of the amplifier. This will allow for the direct retrieval of reasoning results when designing the same amplifier in the future, thereby reducing the consumption of LLMs tokens and improving efficiency.

\subsection{Literature Analysis Agent}
The Literature Analysis Agent is an LLM-based agent equipped with a Retrieval-Augmented Generation (RAG) module and meticulously crafted prompts. These two components work together to ensure that all critical information necessary for amplifier design can be accurately extracted from the under-designed literature, making it directly usable by downstream agents.

\subsubsection{Retrieval-Augmented Generation}
We employ RAG\cite{lewis2020retrieval} to address LLMs' knowledge gaps in multi-stage amplifiers and mitigate the potential impact of hallucination \cite{10.1145/3571730}. 
To be specific, we first use the Mathpix\cite{Mathpix} to convert the under-designed literature from PDF format to Markdown format, and accurately extract the formulas from the literature in latex format.
Next, the converted document which contains the latex-formatted formulas will be embedded as the vector database of the RAG.
These data form the basis of our RAG module.

\begin{figure}[!t]
\centering 
\includegraphics[width=3.4in]{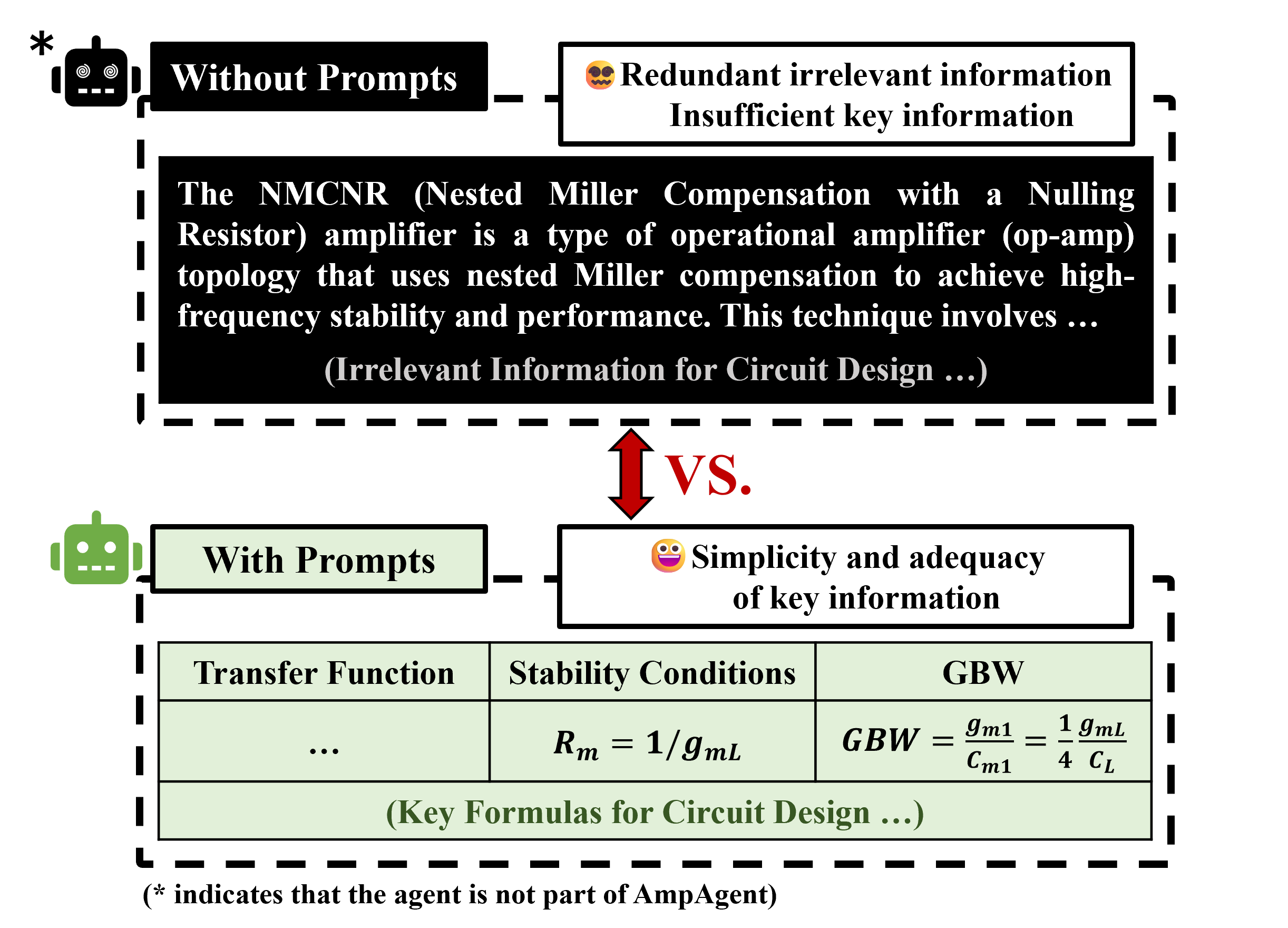}\caption{The comparison between the RAG responses w/ or w/o prompts} 
\label{Fig3.2} 
\end{figure}

Nevertheless, the direct deployment of RAG for circuit design still faces significant challenges in abstracting precise and accurate information. The key information LLM fails to reply despite equipped with the RAG mainly includes 2 parts:
\begin{itemize}
    \item Circuit Schematic: The schematic provided in the literature as an image cannot be directly used in circuit design by LLMs and must first be converted into a netlist format for further parameter sizing. Due to the inaccuracy and inefficiency of existing tools \cite{gurbuz2023img2sim,tao2024amsnet} in handling complex multi-stage amplifier circuits, we are currently converting the schematic images to netlist files manually.
    \item Formulas: The formulas (e.g. transfer function and stability conditions) of an amplifier are frequently dispersed across different sections of literature, which may lead to the loss of critical information and the redundancy of irrelevant information when directly using RAG, as shown in the black box in Fig.\ref{Fig3.2}.
\end{itemize}
Therefore, relying solely on RAG to enhance the LLM's extraction of crucial information about multi-stage amplifiers from literature is still insufficient. Hence, we have further customized a set of prompts to improve the LLM's retrieval capabilities regarding relevant information.

\subsubsection{Prompts}
To address issues of deficiency and redundancy in the retrieval of information by the RAG module for multi-stage amplifier design, we have developed specific prompts for the LLM. By directing the LLM towards targeted knowledge, these prompts enhance the LLM's ability to retrieve the precise information needed for the design process and avoid unnecessary content, making the response succinct and clear.

Fig.\ref{Fig3.2} presents a comparative analysis of the effectiveness of providing prompts versus not providing prompts, focusing on the Nested Miller Compensation with a Nulling Resistor (NMCNR) amplifier as the experimental subject. Fig.\ref{Fig3.2} demonstrates the impact of utilizing prompts on the retrieval process, highlighting the improved accuracy and efficiency achieved when the LLM is guided by specific prompts.

In summary, we integrate the RAG technique with the LLM Agent to efficiently retrieve critical information essential for the multi-stage amplifier design process. By streamlining retrieval and focusing on key data, this approach significantly enhances the efficiency and accuracy of the information retrieval system. However, the retrieved raw information, such as the transfer function, still requires further processing, which necessitates a Mathematics Reasoning Agent to address it.

\begin{figure}[!t]
\centering 
\includegraphics[width=3.4in]{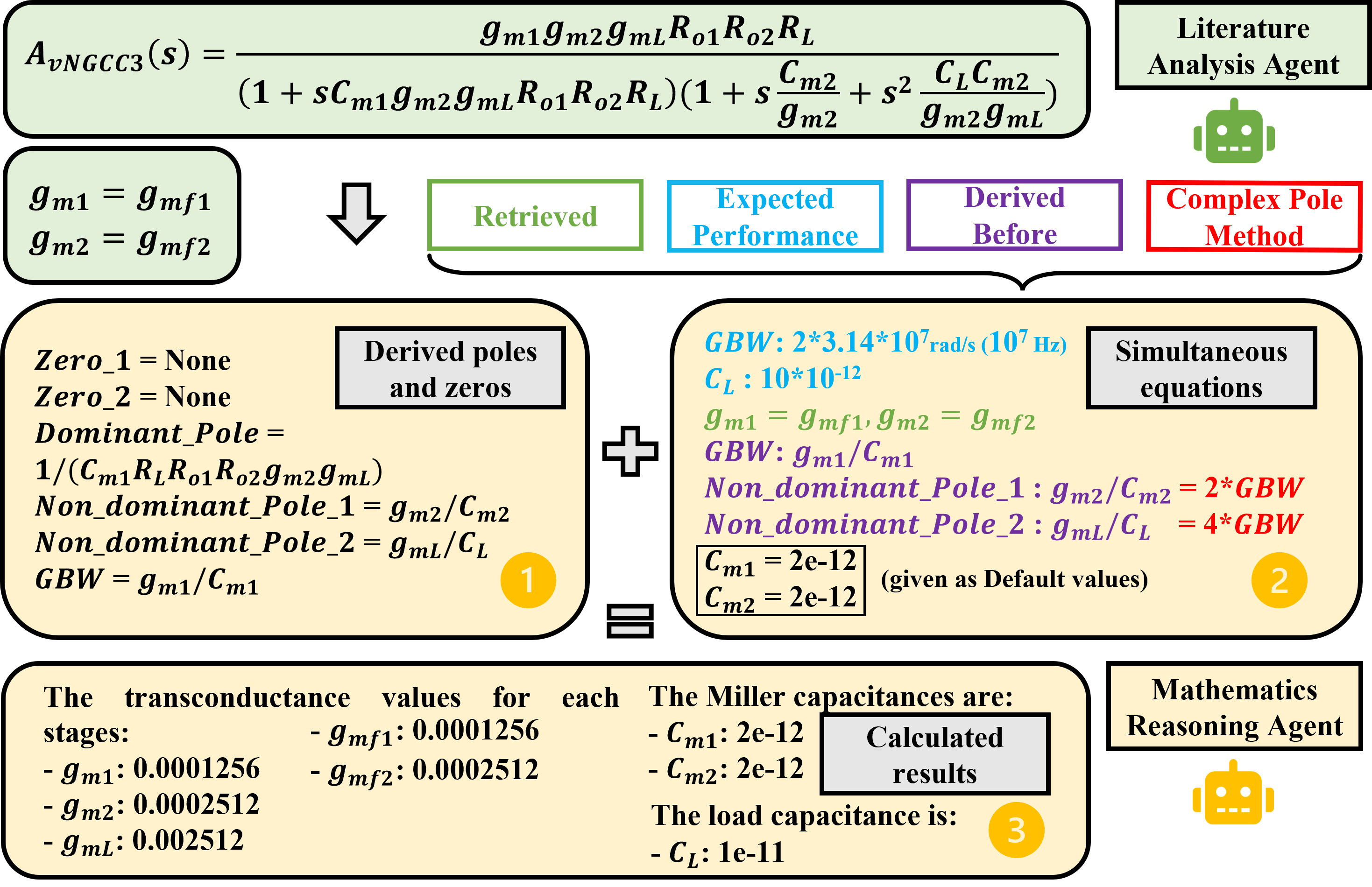}\caption{The response of the Mathematics Reasoning Agent.} 
\label{Fig3.3} 
\end{figure}

\subsection{Mathematics Reasoning Agent}
 Much of the necessary information for the design process has to be derived from the original amplifier's transfer function or calculated through pole-zero distribution methods. In this part, we will illustrate the Mathematics Reasoning Agent dedicated for essential design information derivation.

The module necessitates the sequential implementation of two tasks: (1) the computation of poles and zeros definition based on the transfer function, (2) and the calculation of the targets for the sub-design tasks, taking into account the expected performance metrics, derived poles and zeros' definitions, based on the complex-pole method\cite{eschauzier1995frequency}. For efficient computation, the module is specifically designed to generate Python code to represent the equations that need to be solved and then perform the necessary calculations.

In terms of mathematical reasoning, for tasks with fixed computational paradigms such as calculating pole-zero distribution, we employ a few-shots prompt to convey the rules of the complex-pole method \cite{eschauzier1995frequency} to the Mathematics Reasoning Agent, leveraging its reasoning capabilities to achieve results. For more complex computations, such as solving systems of equations, we instruct the Mathematics Reasoning Agent to utilize the Python SymPy library to ensure accuracy and efficiency.
Here, we give an example of the design derivation of the Nested Gm-C Compensation (NGCC) amplifier \cite{you1997multistage}, when the expected $GBW$ is 10MHz and $C_L$ is 10pF, and the responses of the Mathematics Reasoning Agent is shown in Fig.\ref{Fig3.3}.

As depicted in Fig.\ref{Fig3.3}, the agent has effectively derived the poles and zeros of the amplifier and accurately computed the final $g_m$ for each stage.
This demonstrates the capability of the agent, which is not only able to analyze the design of multi-stage amplifiers but can also derive the $g_m$ and capacitance values when varying input performance metrics.

\subsection{Device Sizing Agent}
\begin{figure}[!t]
\centering 
\includegraphics[width=3.4in]{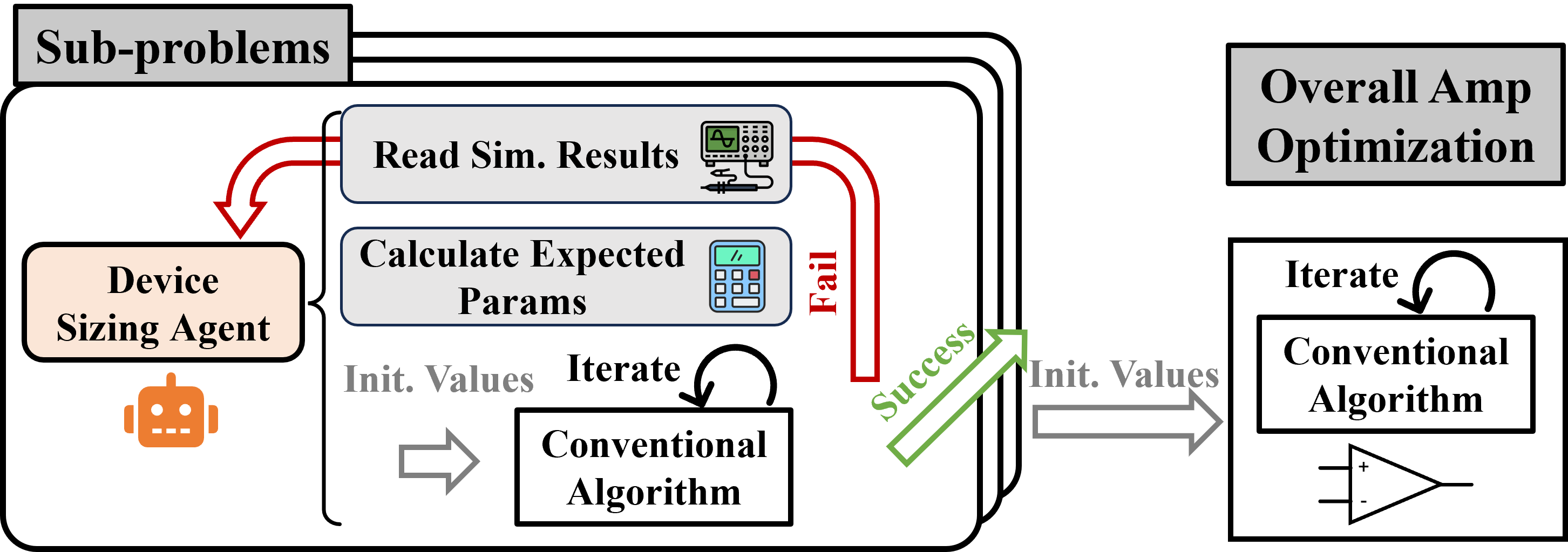}\caption{The overview of the Device Sizing Agent} 
\label{FigCDA} 
\end{figure}

After determining the theoretical design targets for the sub-tasks by the afore agents, specifically the $g_m$ for each stage in the case of an multi-stage amplifier, the Device Sizing Agent is employed. This module is responsible for sizing the transistors and adjusting the device values to attain the desired sub-tasks design targets. The overview of the Device Sizing Agentis shown in Fig.\ref{FigCDA}.

The Device Sizing Agent performs three actions: simulation result reading, accessing calculator, and invoking the conventional optimization algorithms. These actions enable the Device Sizing Agent to interact with the simulator to retrieve current device performance metrics (such as the $g_m$ of transistors, the Gain-Bandwidth product ($GBW$) and Phase Margin ($PM$) of amplifiers, etc.), use the calculator for the estimation of expected device parameters (such as estimate the transistor $W/L$ according to the current $g_m$ and the expected $g_m$), and ultimately use these parameters as initial values for optimization using conventional algorithms such as Artificial Bee Colony (ABC) \cite{karaboga2014comprehensive} and Trust Region Bayesian Optimization (TuRBO) \cite{NEURIPS2019_6c990b7a} algorithm. If the results of the conventional optimization algorithm do not meet the expected metrics, the current device parameters and metrics will be returned to the Device Sizing Agent for it to further estimate the initial values for another loop of optimization.

The process mentioned above will be iteratively repeated until the Device Sizing Agent has optimized all sub-problems. After all sub-problems have been optimized, current device parameters will be used as initial values for overall optimization by the conventional optimization algorithm.

By combining agent-based sizing of the circuit with conventional optimization algorithms, we mitigate the slow iteration rate of large language models (LLMs) and reduce the extensive search space of circuit parameters for the optimization algorithm. This hybrid approach leverages the strengths of both methods, resulting in more efficient and cost-effective circuit design.
The comparative experiments assessing whether to employ optimization algorithms within the sub-problems, as well as the decision to use them in the overall final process, will be shown in Section IV.

\section{Experiment}

In this section, we present the experimental results of seven different types of AmpAgent-designed multi-stage amplifiers. We also compare the performance of AmpAgent with and without conventional optimization algorithms. Furthermore, we evaluate the Device Sizing Agent against conventional optimization algorithms, focusing on the number of algorithm iterations, time consumption, and success rate during parameter sizing. As an example, we will detail the design process for a three-stage Nested Miller Compensation with a Nulling Resistor (NMCNR) amplifier \cite{948432}.

All experiments were conducted on a Linux workstation with one 12th Gen Intel$^\circledR$ Core$^\text{TM}$ i7-12700 CPU and 64-GB memory. The amplifier’s simulations were based on Cadence Spectre$^\circledR$ simulator. The entire AmpAgent framework is developed based on LangChain \cite{langchain}. The LLM we choose is \textit{GPT-4-1106-preview} and its temperature is set as 0. All the AmpAgent-designed amplifiers have been verified through simulation to ensure that no right-half-plane poles are present and that there are no stability issues.

\subsection{NMCNR amplifier}
In this part, we employed AmpAgent in the schematic design of an NMCNR amplifier on the 0.18$\mu m$ process. NMCNR amplifiers \cite{948432} are a kind of 3-stage amplifier with high open-loop gain and a nulling resistor to compensate the right-half-plane zero for a more stable structure \cite{leung1999right}.

\begin{figure}
\centering
    \includegraphics[width=3.6in]{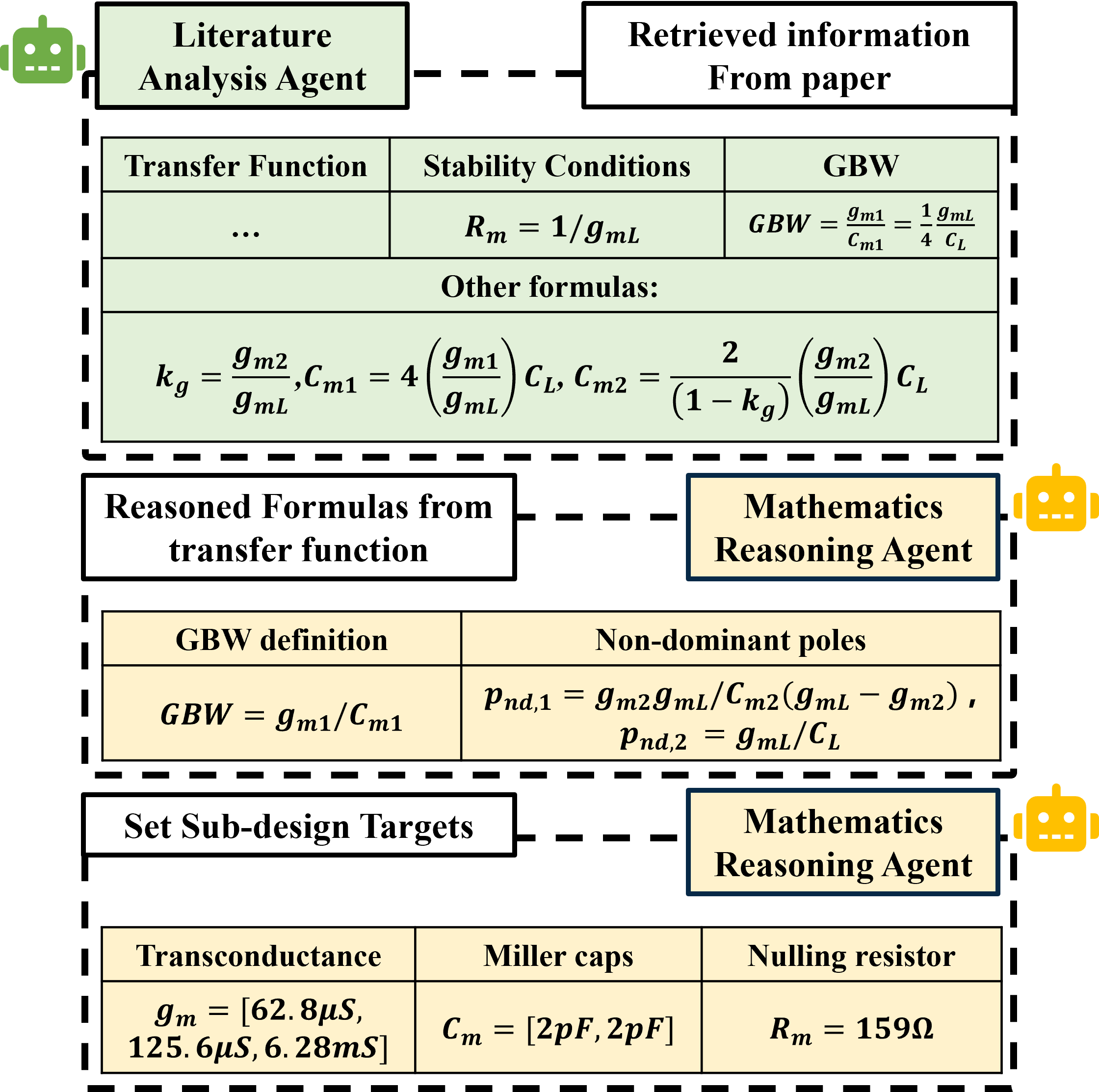}
    \caption{AmpAgent designing a 3-stage NMCNR amplifier with $C_L=50pF$, $GBW = 5MHz$}
    \label{Fig12}
\end{figure}

The literature we selected as the foundation for the AmpAgent to design the NMCMR amplifier is \cite{948432}.
The performance metrics we aimed to achieve by AmpAgent include: $C_L=50pF$, $GBW\geq{5MHz}$, $PM\geq{60^{\circ}}$, $Gain_{DC}\geq{120dB}$, and the power consumption should be as low as possible. The outputs of the Literature Analysis Agent, Mathematics Reasoning Agent are shown in Fig.\ref{Fig12}. The results are shown in Table \ref{tab3}. 

\renewcommand{\arraystretch}{0.5}
\begin{table*}[t]
\scriptsize
\centering
\caption{AmpAgent-designed NMCNR, SMC, NGCC, DFCFC, TCFC, IAC, and AZC amplifiers' performance}
\label{tab3}
\resizebox{\textwidth}{!}{

\begin{tabular}{|c|c|c|c|c|c|c|c|c|c|c|c|c|c|}
\hline
\multirow{2}{*}{Amplifier} &\multicolumn{3}{c|}{Expected Value} &\multicolumn{5}{c|}{w/o further optimization} & \multicolumn{5}{c|}{w/ further optimization} \\
\cline{2-14}
 & \makecell{$GBW$\\$MHz$} & \makecell{$PM$\\$^{\circ}$} & \makecell{$Gain_{DC}$\\${dB}$} & 
\makecell{$GBW$\\$MHz$} & \makecell{$PM$\\$^{\circ}$} & \makecell{$Gain_{DC}$\\${dB}$} & \makecell{$I_{dd}$\\${mA}$} & \makecell{$IFOM_S$\\ \tiny{$\frac{MHz*pF}{mA}$}} &
\makecell{$GBW$\\$MHz$} & \makecell{$PM$\\$^{\circ}$} & \makecell{$Gain_{DC}$\\${dB}$} & \makecell{$I_{dd}$\\${mA}$} & \makecell{$IFOM_S$\\ \tiny{$\frac{MHz*pF}{mA}$}}\\
\hline
\makecell{NMCNR \cite{948432}\\@$C_L=50pF$ } & $\geq5.00$ & $\geq60$ & $\geq120$ & $5.15$ & $32.5$ & $135$ & $0.38$ & 675 & $5.01$ & $61.5$ & $131$ & $0.322$ & \textbf{778} \\
\hline
\makecell{SMC \cite{allen2011cmos}\\@$C_L=10pF$ } & $\geq10.00$ & $\geq60$ & $\geq70$ & $9.47$ & $60.05$ & $73.53$ & $0.087$ & 1092 & $10.0$ & $60.7$ & $70.06$ & $0.080$ & \textbf{1243}\\
\hline
\makecell{NGCC \cite{you1997multistage}\\@$C_L=50pF$ } & $\geq1.00$ & $\geq60$ & $\geq100$ & $1.24$ & $52$ & $137$ & $0.078$ & 795 & $1.06$ & $63$ & $136$ & $0.054$ & \textbf{981}\\
\hline
\makecell{DFCFC \cite{leung1999damping}\\@$C_L=150pF$ } & $\geq2.00$ & $\geq60$ & $\geq100$ & $2.00$ & $1.2$ & $119$ & $0.035$ & \textbf{8645} & $2.12$ & $60$ & $100$ & $0.052$ & 6103\\
\hline
\makecell{TCFC  \cite{peng2005transconductance}\\@$C_L=300pF$ } & $\geq2.00$ & $\geq60$ & $\geq100$ & $2.05$ & $77$ & $127$ & $0.053$ & 11647 & $2.28$ & $71$ & $130$ & $0.017$ & \textbf{41204}\\
\hline
\makecell{IAC  \cite{peng2010impedance}\\@$C_L=500pF$ } & $\geq4.00$ & $\geq60$ & $\geq100$ & $3.55$ & $57$ & $153$ & $0.053$ & 33490 & $4.87$ & $76$ & $150$ & $0.039$ & \textbf{62435}\\
\hline
\makecell{AZC  \cite{yan20130}\\@$C_L=15nF$ } & $\geq1.00$ & $\geq60$ & $\geq100$ & $0.96$ & $58$ & $135$ & $0.049$ & 195918 & $1.01$ & $61.0$ & $133$ & $0.047$ & \textbf{322340}\\
\hline
\end{tabular}
}
\end{table*}
\renewcommand{\arraystretch}{1}
\begin{table}[t]
    \begin{center}
    \caption{AmpAgent-designed Amplifiers' and Original literature's $IFOM_S$ comparison}
    \label{improvement}

    \begin{tabular}{|c|c|c|c|}
    \hline
    Amplifier Name & \makecell{$IFOM_S$\\ \tiny{$\frac{MHz*pF}{mA}$} \\ Origin literature} & \makecell{$IFOM_S$ \\ \tiny{$\frac{MHz*pF}{mA}$}\\ By AmpAgent} & Improvements \\
    \hline
    SMC\cite{allen2011cmos} & 400 & 1243 & 3.11$\times$\\
    \hline
    NMCNR\cite{948432} & 410 & 778 & 1.90$\times$ \\
    \hline
    NGCC\cite{you1997multistage} & 36 & 981 & \bm{$27.25\times$}\\
    \hline
    DFCFC\cite{leung1999damping} & 1238 & 6103& 4.93$\times$\\
    \hline
    TCFC\cite{peng2005transconductance} & 14250 & 41204& 2.89$\times$\\
    \hline
    IAC\cite{peng2010impedance} & 33000 & 62435& 1.89$\times$\\
    \hline
    AZC\cite{yan20130} & 197916 & 322340& \bm{$1.63\times$}\\
    \hline
    \end{tabular}
    \end{center}
\end{table}

\subsection{Other multi-stage amplifiers}
Except for the amplifiers mentioned above, we also test AmpAgent's capacity in design more types of multi-stage amplifiers. In this part, we will present the design results of AmpAgent for several amplifiers with typical compensation structures, including Single Miller Compensation (SMC) \cite{allen2011cmos}, NGCC \cite{you1997multistage}, Damping-Factor-Control Frequency-Compensation (DFCFC) \cite{leung1999damping}, Transconductance with Capacitor Feedback Compensation (TCFC) \cite{peng2005transconductance}, Impedance Adapting Compensation (IAC) \cite{peng2010impedance}, and Active Zero Compensation (AZC) amplifiers \cite{yan20130} across different requirements for $GBW$ and $C_L$.

Additionally, we compared the amplifiers' performance with and without the implementation of overall optimization following the Device Sizing Agent. The experiment results are shown in Table \ref{tab3}.
From Table \ref{tab3}, we can observe that incorporating overall optimization allows the amplifiers to meet the constraints, and further enhances their performance. 
We utilize the $IFOM_S$, which is defined in (\ref{IFOMS}), to evaluate the amplifier performance. $IFOM_S$ is a common metric in literature \cite{peng2010impedance,peng2005transconductance,yan20130}.
\begin{equation}
    IFOM_S = GBW*C_L/I_{dd}\label{IFOMS}
\end{equation}
The improvement in various amplifiers, compared to the original literature's $IFOM_S$, are also listed in Table \ref{improvement}. The performance for these amplifiers have seen an enhancement ranging from 1.63${\times}$ to 27.25${\times}$ compared to original paper.

\begin{table*}[htbp]
\renewcommand{\arraystretch}{1}
\begin{center}
\caption{Comparison between This Work, ABC, and TuRBO in terms of iteration number, time consumption, and success rate}
\label{tab4}
\begin{tabular}{|c|c|c|c|c|c|c|c|c|c|c|c|c|}
\hline 
\multirow{2}{*}{Amplifier} & \multicolumn{3}{c|}{This Work} & \multicolumn{3}{c|}{ABC \cite{karaboga2014comprehensive}} & \multicolumn{3}{c|}{TuRBO-5 \cite{NEURIPS2019_6c990b7a}} & \multicolumn{3}{c|}{TuRBO-1 \cite{NEURIPS2019_6c990b7a}}\\
\cline{2-13}
 & \makecell{Avg. \\ \# Iter.} & \makecell{Avg.\\ Time (s)} & \makecell{Success\\rate} & \makecell{Avg. \\ \# Iter.} & \makecell{Avg.\\ Time (s)} & \makecell{Success\\rate} & \makecell{Avg. \\ \# Iter.} & \makecell{Avg.\\ Time (s)} & \makecell{Success\\rate} & \makecell{Avg. \\ \# Iter.} & \makecell{Avg.\\ Time (s)} & \makecell{Success\\rate}\\
\hline
\makecell{SMC \cite{allen2011cmos}}& 16 & 245 & 100\% & 35 & 456 & 68\% & 38 & 460 & 95\% & 56 & 540 & 76\% \\
\hline
\makecell{NMCNR \cite{leung1999right}}& 19 & 444 & 100\% & 75 & 600 & 46\% & 53 & 842 & 84\% & 52 & 597 & 82\% \\
\hline
\makecell{NGCC \cite{you1997multistage}}& 22 & 458 & 98\% & 75 & 543 & 46\% & 29 & 1200 & 95\% & 35 & 843 & 94\% \\
\hline
\makecell{DFCFC \cite{leung1999damping}}& 33 & 603 & 95\% & 93 & 783 & 14\% & 53 & 1321 & 83\% & 46 & 836 & 88\% \\
\hline
\makecell{TCFC \cite{peng2005transconductance}}& 26 & 471 & 96\% & 75 & 601 & 54\% & 36 & 894 & 89\% & 48 & 768 & 85\% \\
\hline
\makecell{IAC \cite{peng2010impedance}}& 25 & 481 & 97\% & 100 & 1352 & (Failed) & 61 & 976 & 73\% & 64 & 742 & 82\% \\
\hline
\makecell{AZC \cite{yan20130}}& 38 & 696 & 96\% & 100 & 1465 & (Failed) & 100 & 2083 & (Failed) & 100 & 1610 & (Failed) \\
\hline
\end{tabular}
\end{center}
\end{table*}

\subsection{Integration with conventional optimization algorithms}
In addition to using optimization algorithm for the overall optimization of the amplifier, AmpAgent can also employ the algorithm to assist in sizing the transistors during the resolution of sub-problems, as discussed in Section III.C.

In this part, we will compare whether AmpAgent uses an optimization algorithm, specifically ABC, during the optimization process, and we will look at the differences in terms of LLM token consumption and design time.
\begin{figure}[!t]
\centering
  \includegraphics[width=3in]{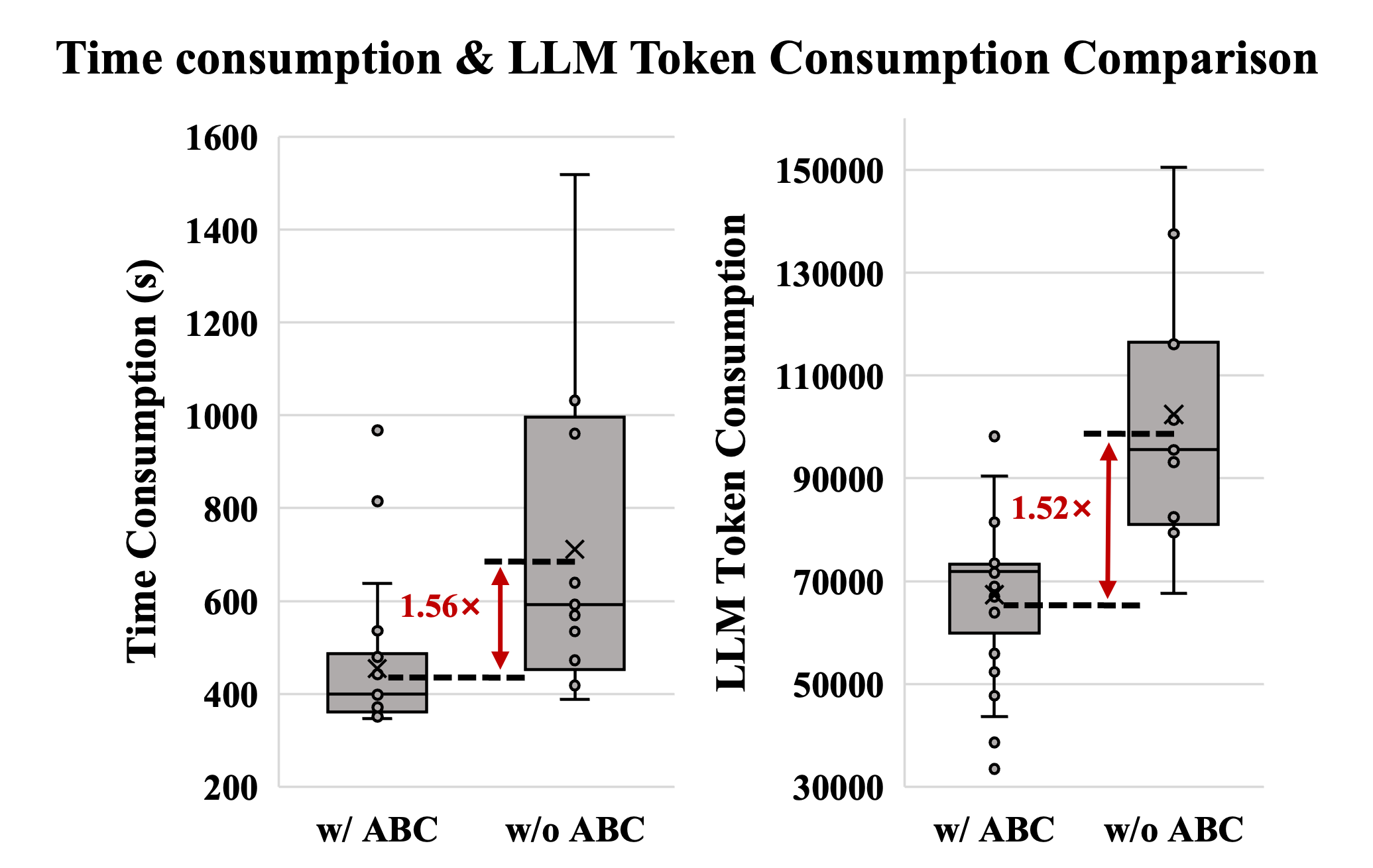}
  \caption{Design time and token consumption comparison} 
  \label{Fig13}
\end{figure}
Taking the automatic design of the NGCC \cite{you1997multistage} amplifier by AmpAgent as an example, we randomly sample the expected values of $GBW$ and $C_L$ within the range of $1MHz$ to $10MHz$ and $10pF$ to $100pF$.
Comparisons of time consumption and token consumption by the LLM, here GPT4, are shown in Fig. \ref{Fig13}. Fig. \ref{Fig13} shows that AmpAgent, by adopting optimization algorithm to tackle sub-optimization issues, has reduced the time required by 1.56${\times}$ and improved token consumption efficiency by 1.52${\times}$ compared to not using it.
From this comparison, it can be observed that AmpAgent is better suited for decomposing sub-optimization problems and providing initial solutions for these sub-problems, rather than directly performing iterative optimization on parameters itself.

\subsection{Optimization Efficiency}
Furthermore, we conducted a comparative analysis between the Device Sizing Agent and the conventional optimization algorithms in terms of efficiency in parameter sizing. This comparison, which is shown in Table \ref{tab4}, focused on the number of iterations, design time, and success rate. We set the maximum number of iterations to 100. The results convincingly demonstrate that AmpAgent can significantly enhance the accuracy of automated design processes while effectively reducing the number of iterations and the associated design time. It is noteworthy that the instances of design failure associated with AmpAgent can be attributed to the occasional hallucinatory effects of GPT-4, which lead to its inability to execute certain actions correctly.

\section{Conclusion}
In this study, we presented AmpAgent, which is a novel approach utilizing an LLM-based multi-agent dedicated to multi-stage schematic design from literature with process and performance porting. AmpAgent consists of three agents, and through a detailed division of design problem, it mitigates the current issue of LLMs lacking in specialized knowledge and design process understanding for analog circuits.

We specifically showcase AmpAgent's effectiveness in auto design of seven different types of multi-stage amplifiers with process and performance porting.
The performance for these amplifiers has seen an enhancement ranging from 1.63${\times}$ to 27.25${\times}$ compared to the original literature, confirming AmpAgent's capability of such complex analog circuits. 
The design efficiency of AmpAgent surpasses that of conventional optimization algorithms, achieving a significant reduction in iteration counts by 1.32$ \sim $4${\times}$ and execution time by 1.19$ \sim $2.99${\times}$. It also boasts an enhanced success rate by 1.03$ \sim $6.79${\times}$ and it can even successfully design amplifiers where conventional optimization algorithms fail.

To further enhance AmpAgent, future work could involve training an analog-circuit-domain LLM, automation of netlist extraction from circuit diagram and integration of advanced circuit design methodologies like $g_m/I_d$\cite{jespers2009gm} for higher efficiency.

\bibliographystyle{IEEEtran}
\bibliography{AmpAgent-DATE.bib}

\end{document}